\documentclass[epjCONF]{svjour}

\newcommand{\met}{\ensuremath{{\slash\kern-.7emE}_{T}}}

\usepackage{graphics}
\usepackage[varg]{txfonts} 
\usepackage[latin1]{inputenc}
\session-title{XXIInd Hadron Collider Physics Symposium}
\begin{document}
\title{W Mass results from Tevatron and LHC}
\author{Alex Melnitchouk for ATLAS, CDF, CMS, D0 and LHCb Collaborations\fnmsep\thanks{\email{alexmelnitchouk@gmail.com}}}

\institute{University of Mississippi, University, Mississippi, 38677, USA}
\abstract{
Most recent results of $W$ boson mass measurements from Tevatron experiments (CDF and D0) in $p\bar{p}$ collisions at 
$\sqrt{s}=1.96$~GeV are reported, using $0.2~fb^{-1}$ and $1.0~fb^{-1}$ data collected at CDF and D0, respectively. 
The measurements of $W$ boson properties at LHC experiments (ATLAS, CMS, and LHCb) in $pp$ collisions at $\sqrt{s}=7$~TeV, 
using data collected before Summer 2011, are presented. 
These measurements are essential at the preparation stage of the $W$ boson mass measurements at LHC.
Challenges for $W$ mass measurement at the LHC in comparison with the Tevatron are outlined.
Prospects for $W$ mass precision with upcoming measurements and its implications are discussed.
} 
\maketitle
\section{Introduction}
\label{intro}
Measurement of the $W$ boson mass ($M_{W}$) 
provides us with a uniquely powerful key to uncovering the origin of the electroweak
symmetry breaking and learning about new physics.
At the loop level, W boson is connected with the top quark and the Higgs boson
via the radiative corrections to the W mass. 
Hence precise measurements of
the masses of the top quark and W boson allow us to constrain the
most probable mass range of the Higgs boson mass.
Current world average for W mass is 80.399$\pm$0.023 GeV~\cite{comb}. 
Current world average for top quark mass is 173.2$\pm$0.9 GeV~\cite{combtop}.
These measurements combined with other precision measurements tell us that
the mass of the Standard-Model Higgs boson is lower than 161 GeV at 95\% confidence level~\cite{lepwg}.
With improved precision of the W boson mass measurement tighter constraints could be placed 
on the Higgs boson mass. Compatibility of such tighter constraints from precision data with 
the results from ongoing direct Higgs boson searches
or lack thereof would be a critical piece of information for understanding
electroweak symmetry breaking mechanism.

\section{Identification of Electrons and Muons at CDF and D0}
\label{sec:1}
Electrons are identified as an electromagnetic (EM) cluster 
reconstructed with a simple cone algorithm. 
To reduce the background of jets faking electrons, 
electron candidates are required to have a large fraction 
of their energy deposited in the EM section of the calorimeter 
and pass energy isolation and shower shape requirements.
Electron candidates are classified as \textit{tight} 
if a track is matched spatially to EM cluster and if the track 
transverse momentum is close to the transverse energy of the EM cluster.
In CDF~\cite{cdf_det} electrons are reconstructed both in the central
calorimeter and plug calorimeter ($|\eta| < 2.8$) 
while electrons in D0~\cite{d0_det} are reconstructed 
in the central and endcap calorimeters ($|\eta|<1.05$ and $1.5<|\eta|<3.2$). 
Here $\eta = -\ln\tan(\theta/2$, and  $\theta$ is the polar angle with respect 
to the proton direction. Both CDF and D0 require  \textit{tight} electrons 
in the central calorimeter ($|\eta|<1.05$) for $W \rightarrow e\nu$ 
candidates. Electron energies are measured with the calorimeter,
while electron direction is measured with tracking detectors, using
tracks that are matched to electron cluster in the calorimeter.

Muons are identified by a track in the muon system matched to a track 
in the central tracking system.
Measurements include the muons reconstructed in the central muon extension sub-detector which 
extends the coverage from $|\eta|<0.6$ to $|\eta|<1.$

\section{Overview of $W$ mass measurement}
\label{sec:2}
W boson mass is measured using three transverse kinematic variables: 
the transverse mass\newline $m_{T} = \sqrt {2p_{T}^{e,\mu}p_{T}^{\nu}(1 - \cos\Delta\phi)}$, 
the the transverse momentum of the lepton\footnote{electron or muon in the context of
this usage, D0 uses only electron channel, 
whereas CDF uses both electron and muon channels for $M_{W}$ measurement} 
( $p_{T}^{e,\mu}$) and neutrino ($p_{T}^{\nu}$) 
transverse momentum, where $\Delta\phi$ is the opening angle 
between the electron(muon) and neutrino momenta in the plane transverse to the beam.
Neutrino transverse momentum ($p_{T}^{\nu}$) is inferred from the imbalance
of transverse energy. We also refer to this observable as missing $E_{T}$ (MET). 

A sophisticated parametrized Monte Carlo simulation is used 
for modeling these variables as a function of $M_{W}$.
$M_{W}$ is extracted from a binned maximum-likelihood fit between the data and simulation.
Fast simulation includes models of electron(muon), recoil system, and backgrounds.
Electron efficiencies, resolution and energy scale parameterizations are tuned
to $Z \rightarrow ee$ data.

Recoil system represents energy deposited in the calorimeter from all
sources except the electron(s). Recoil system consists of three major components:
hard recoil (particles that collectively balance the $p_{T}$ of the $W$ of $Z$ boson),
underlying event, and additional interactions. Contribution from the third component depends 
on the instantaneous luminosity. Hard recoil is modeled using the full detector simulation,
while the other two components are described by real data events.
Full recoil model is tuned to $Z \rightarrow ee$ data, using imbalance between
the $Z$ boson momentum measured with electrons(muons) and with recoil system.
Sources of backgrounds to $W \rightarrow e,\mu\nu$ events include $W \rightarrow \tau\nu \rightarrow e,\mu\nu\nu$,
QCD, and $Z \rightarrow ee,\mu\mu$ processes.

\section{Lepton Energy Scale Calibration}
\label{sec:3}
Dominant uncertainties in $M_W$ measurements come 
from lepton energy scale measurements. To first order
fractional error on the lepton energy scale translates to fractional error 
on the W mass\cite{ashutoshandjan}. 

D0 determines electron energy scale
using high $p_{T}$ electrons from $Z \rightarrow ee$ decays.
Precision of such calibration is limited mostly by the size of
the $Z \rightarrow ee$ sample.

\begin{figure}[hbpt]
\resizebox{0.95\columnwidth}{!}{
\includegraphics{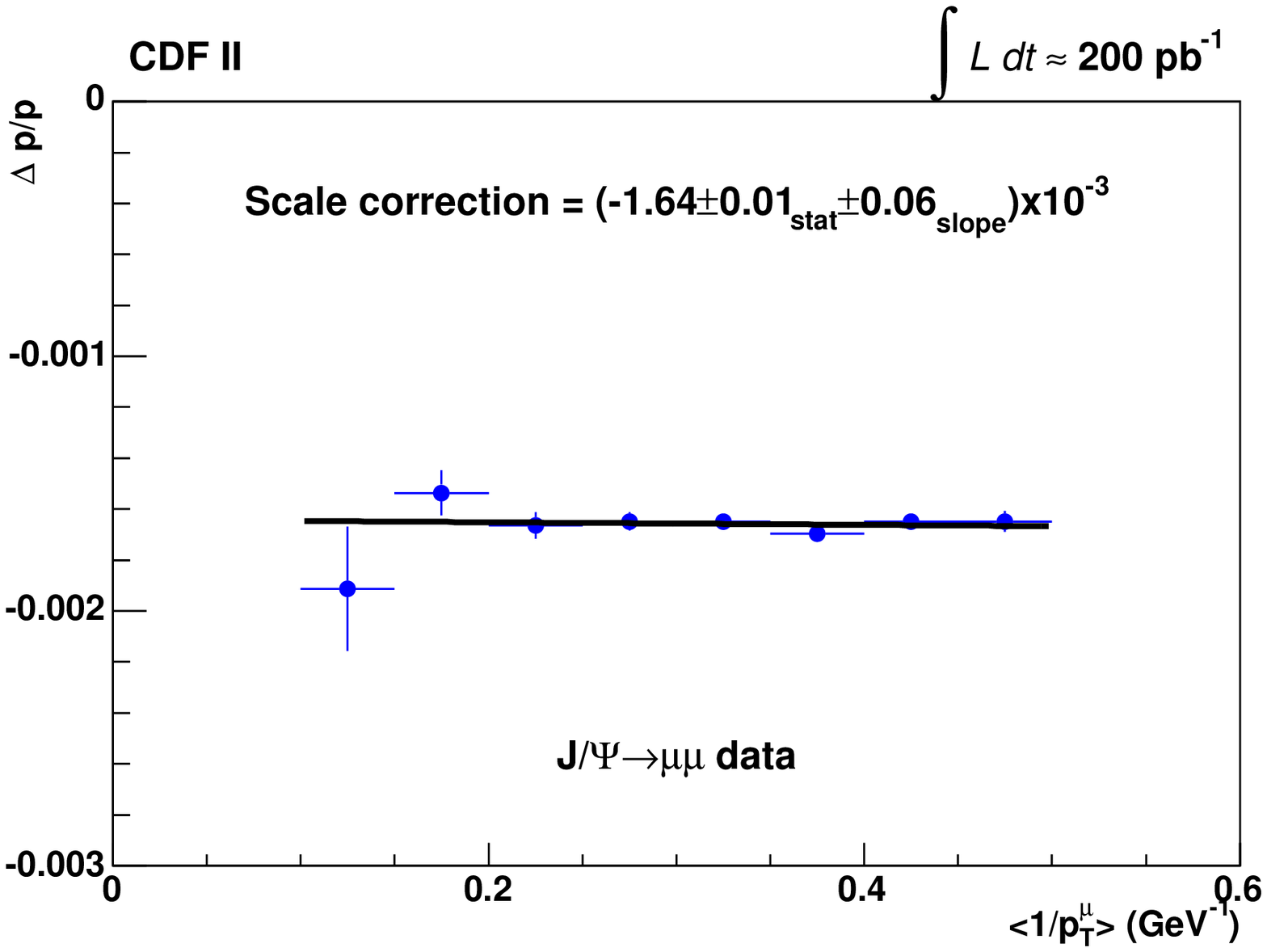}}
\resizebox{0.95\columnwidth}{!}{\includegraphics{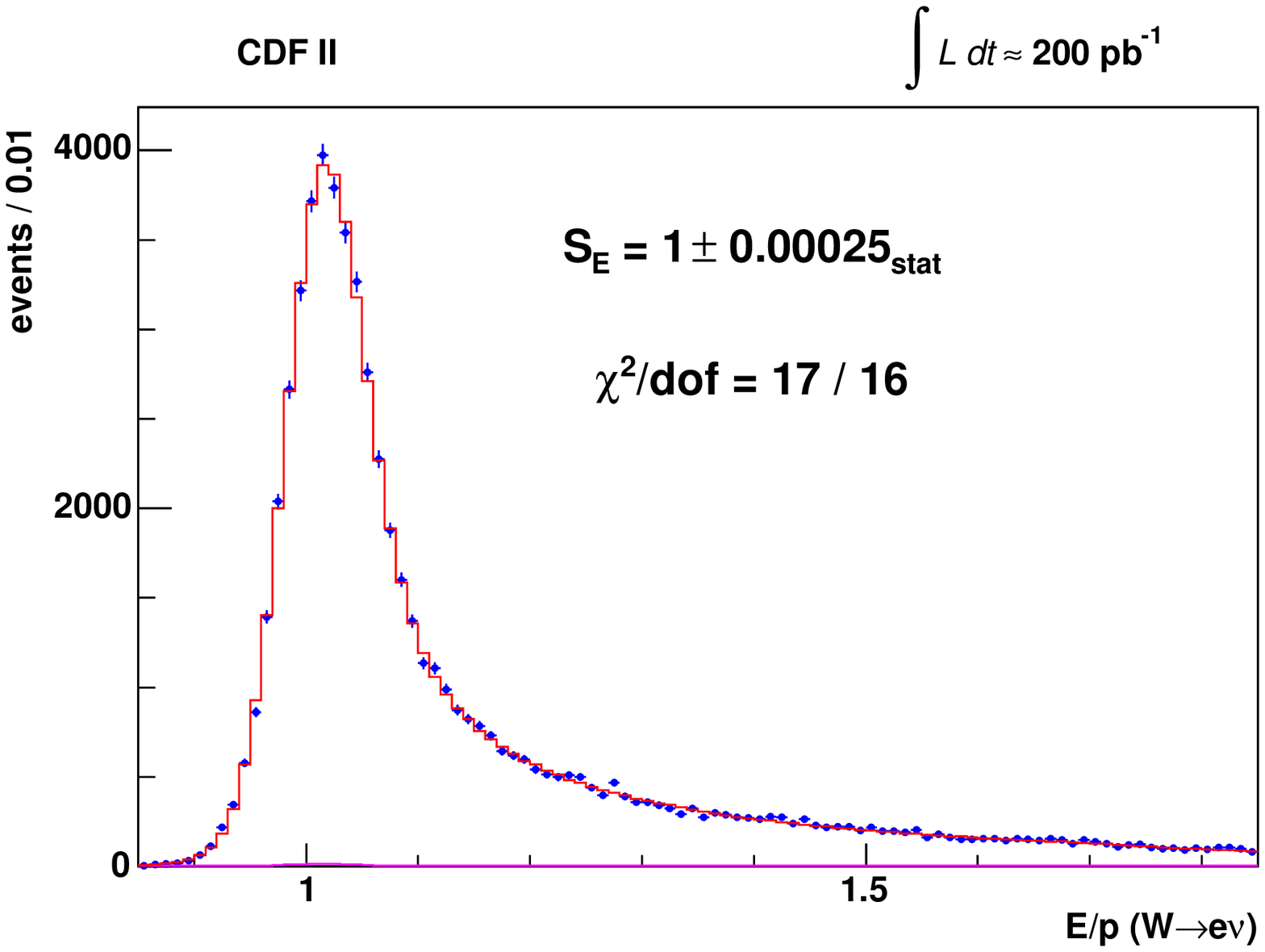}}
\caption{Top: fractional muon momentum correction as a function of inverse momentum.
Bottom: ratio of electron energy measured in the calorimeter
to electron momentum measured by the tracking system in $W \rightarrow e\nu$ events.}
\label{fig:cdfjpsimass}
\end{figure}

CDF relies on tracking detector for both
electron and muon energy scale calibration. First tracking detector
is calibrated using $J/\psi \rightarrow \mu\mu$ events. $J/\psi$
invariant mass is measured as a function of muon momentum.
Fig.~\ref{fig:cdfjpsimass} shows the correction needed
to make measured $J/\psi$ mass to be at its PDG value (overall offset) and 
independent of muon momentum (slope). This correction was implemented
in the simulation by adjusting the energy-loss model. Then tracker calibration
is transported to the calorimeter using $W \rightarrow e\nu$ electrons
near the peak of the E/p distribution, shown also in Fig.~\ref{fig:cdfjpsimass}.

\section{Results and Prospects}
\label{sec:4}
\begin{figure}[hbpt]
\begin{center}
\resizebox{0.99\columnwidth}{!}{  \includegraphics{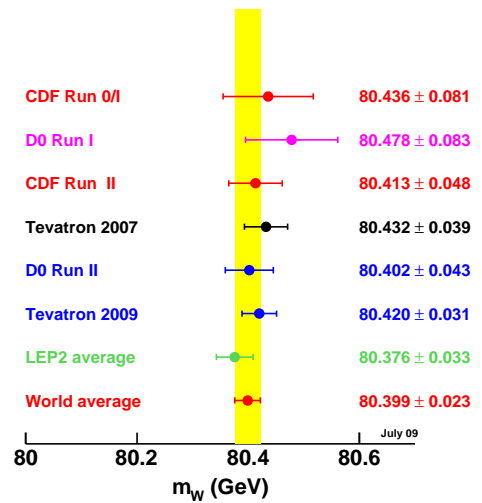} }
 \end{center}
\vspace{-1pc}
  \caption{
Summary of the measurements of the $W$ boson mass and their average. 
The result from the Tevatron corresponds to the values which includes 
corrections to the same W boson width and PDFs. 
The LEP II results are from~\cite{lep}. 
An estimate of the world average of the Tevatron and LEP 
results is made assuming no correlations between the Tevatron and LEP 
uncertainties.
}
\label{fig:mwfig}
\end{figure}

\begin{figure*}[hbpt]
\begin{center}
\resizebox{0.68\columnwidth}{!}{  \includegraphics{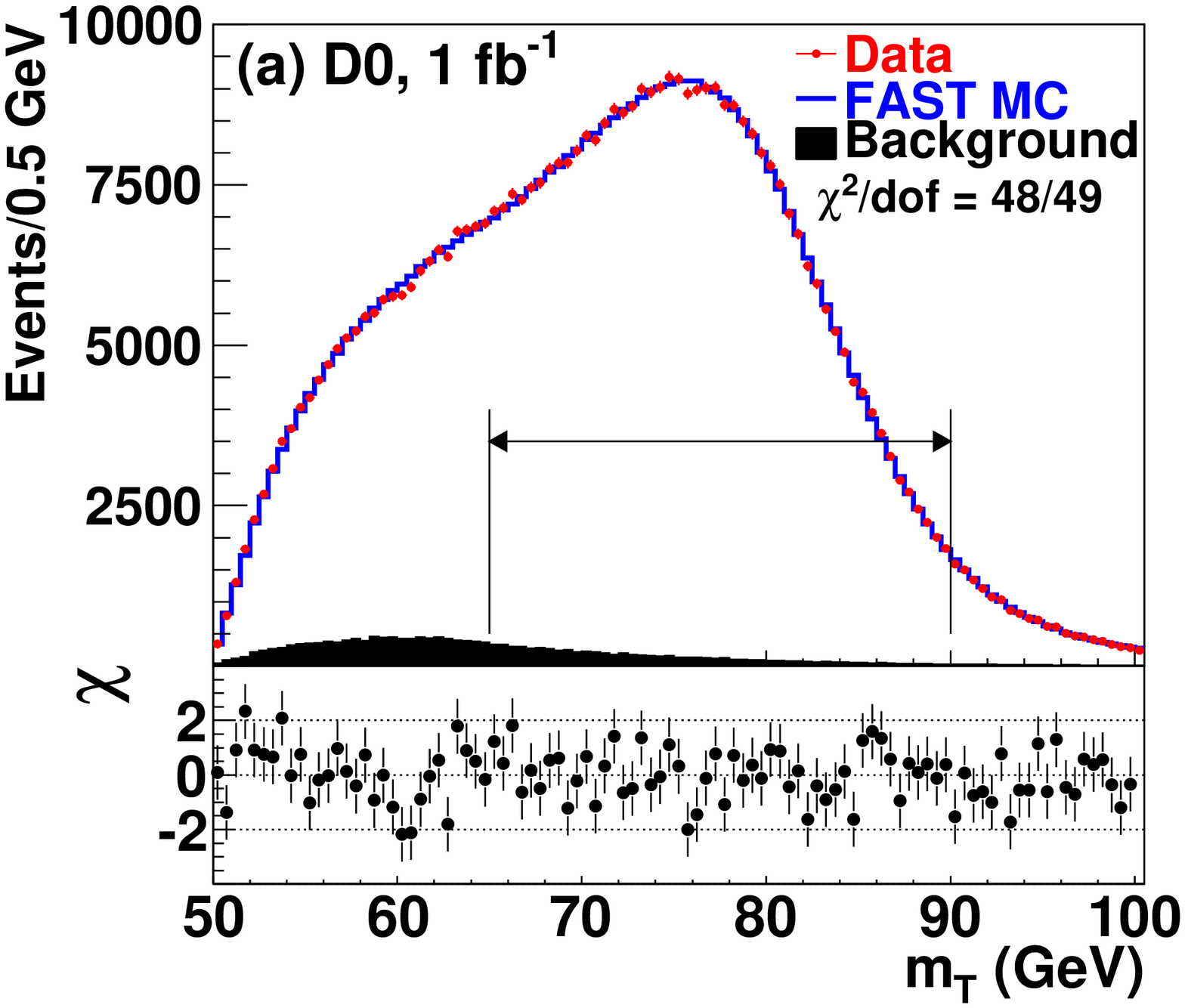}      }
\resizebox{0.68\columnwidth}{!}{  \includegraphics{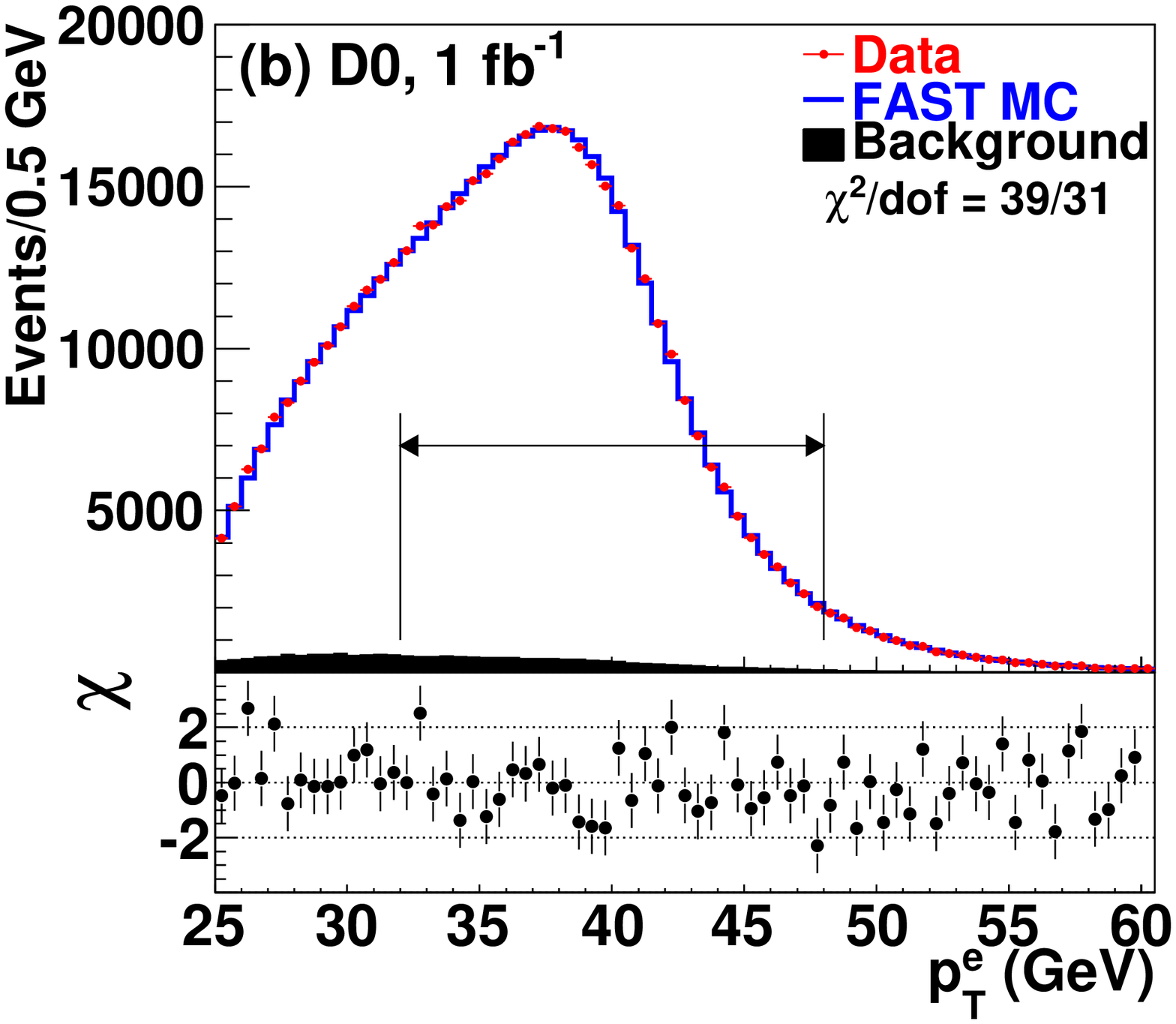}  }
\resizebox{0.68\columnwidth}{!}{  \includegraphics{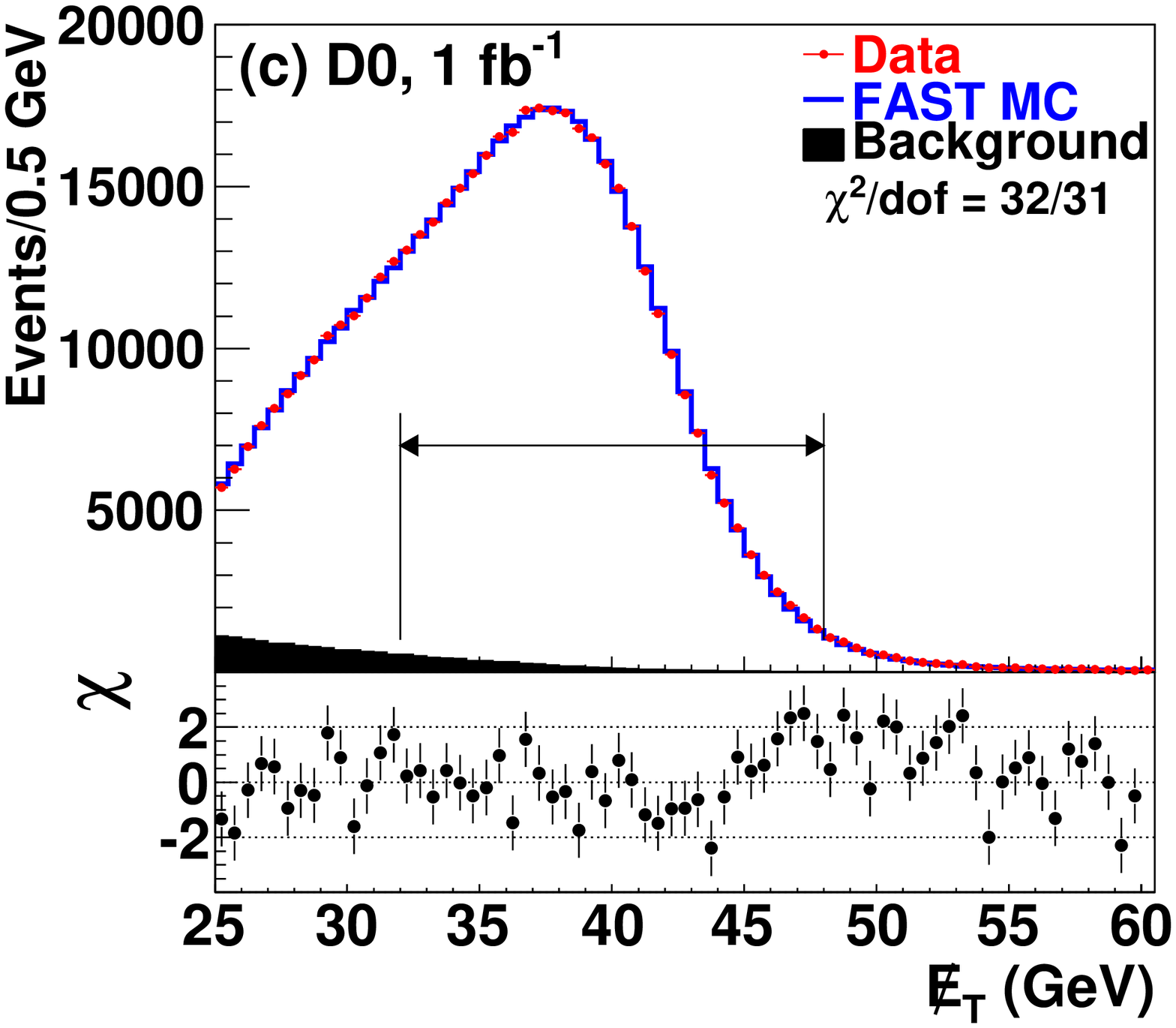}     }
\end{center}
  \caption{
Eelectron $m_{T}$,  $p_T$, and MET distributions in $W \rightarrow e\nu$ D0 data and fast simulation ({\sc fastmc}).
Added background is shown as well. Signed $\chi$ distributions are shown in the bottom of part of each plot.
Signed $\chi$ is defined as
$\chi_i = [N_i-\,(${\sc fastmc}$_i)]/\sigma_i$ 
for each point in the distribution, $N_i$ is
the data yield in bin $i$ and $\sigma_i$ is the 
statistical uncertainty in bin $i$.
}
\label{fig:mwd0}
\end{figure*}

\begin{figure*}[hbpt]
\begin{center}
\resizebox{0.69\columnwidth}{!}{\includegraphics{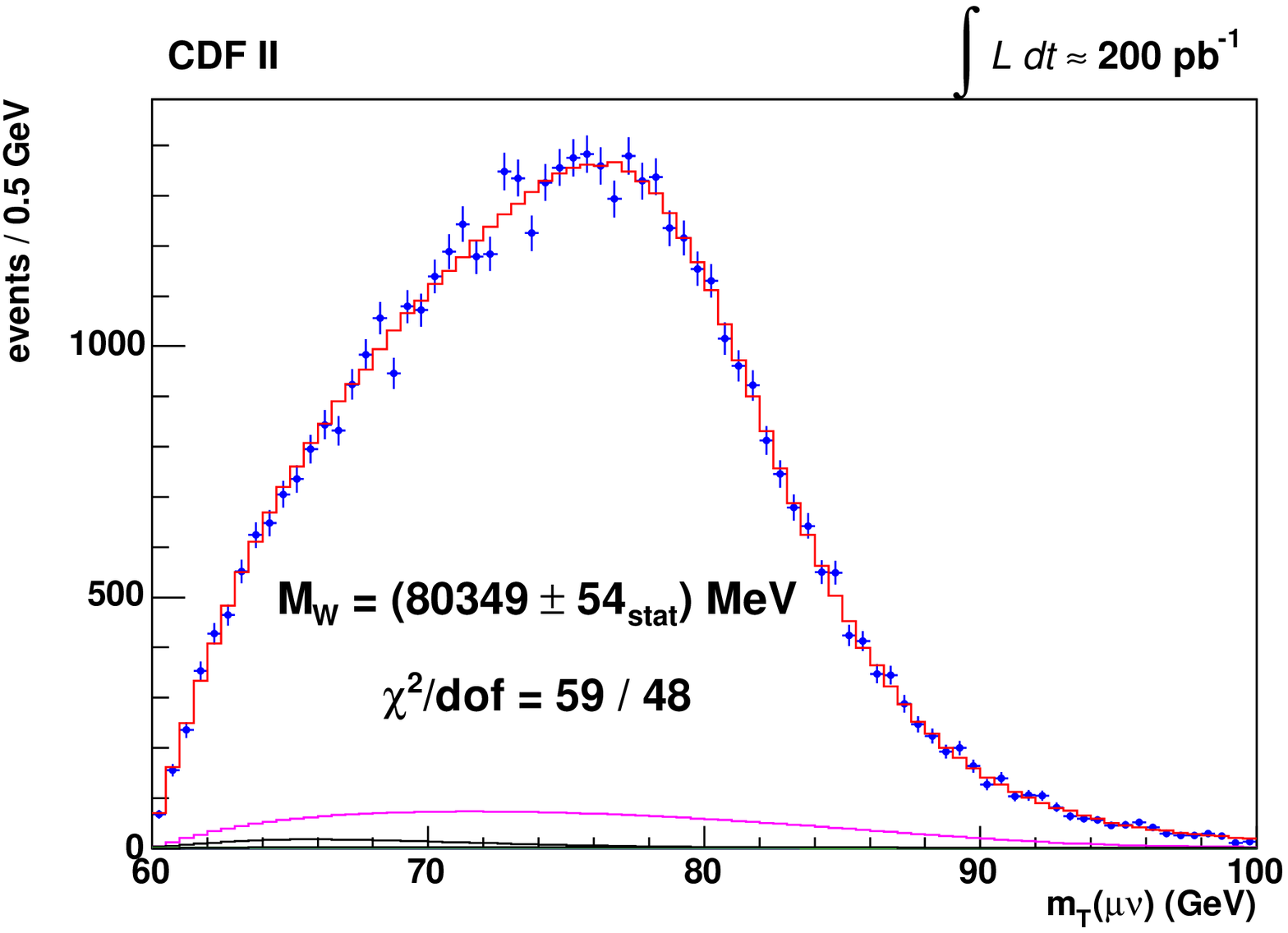}}
\resizebox{0.69\columnwidth}{!}{\includegraphics{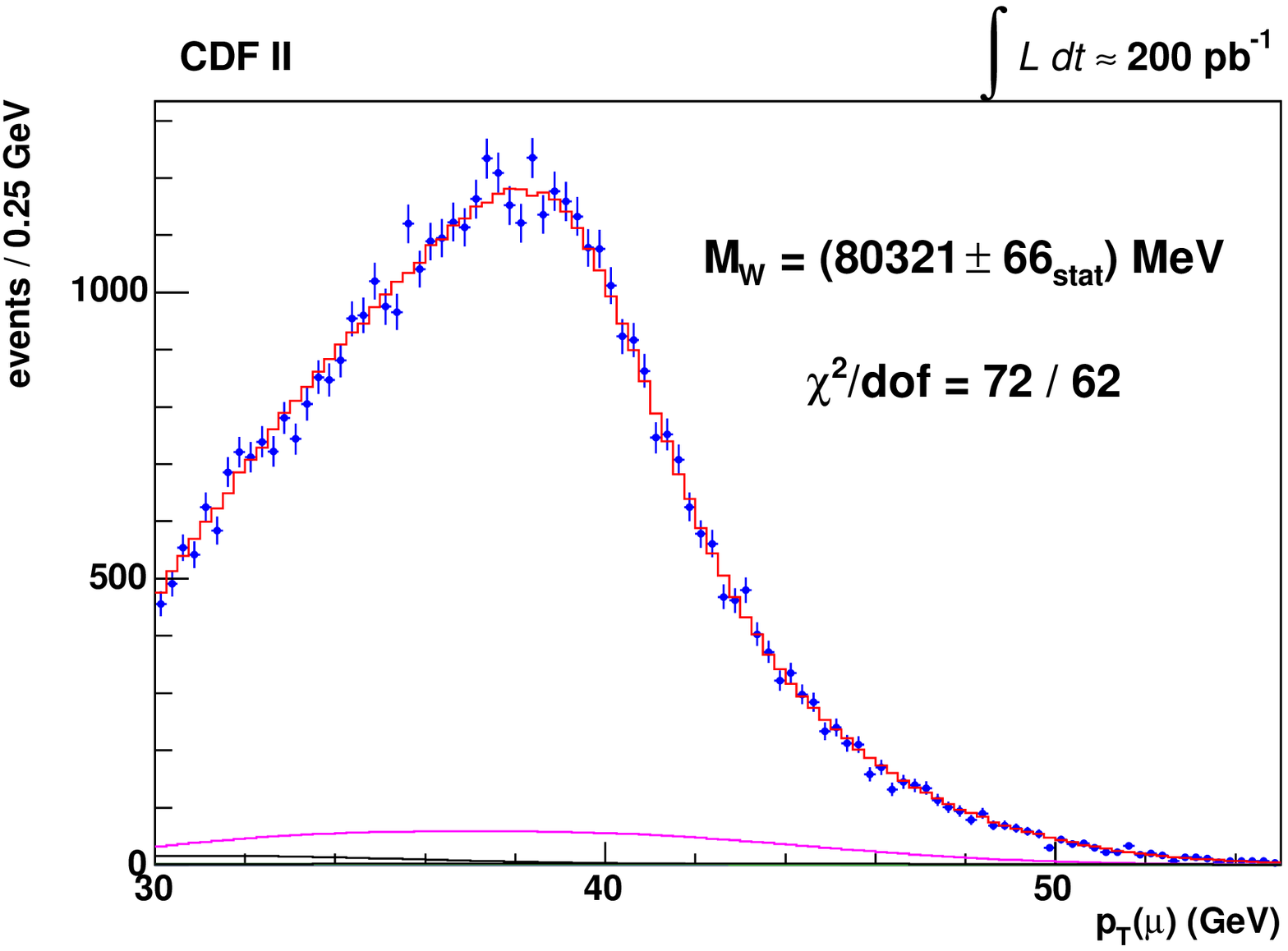}}
\resizebox{0.69\columnwidth}{!}{\includegraphics{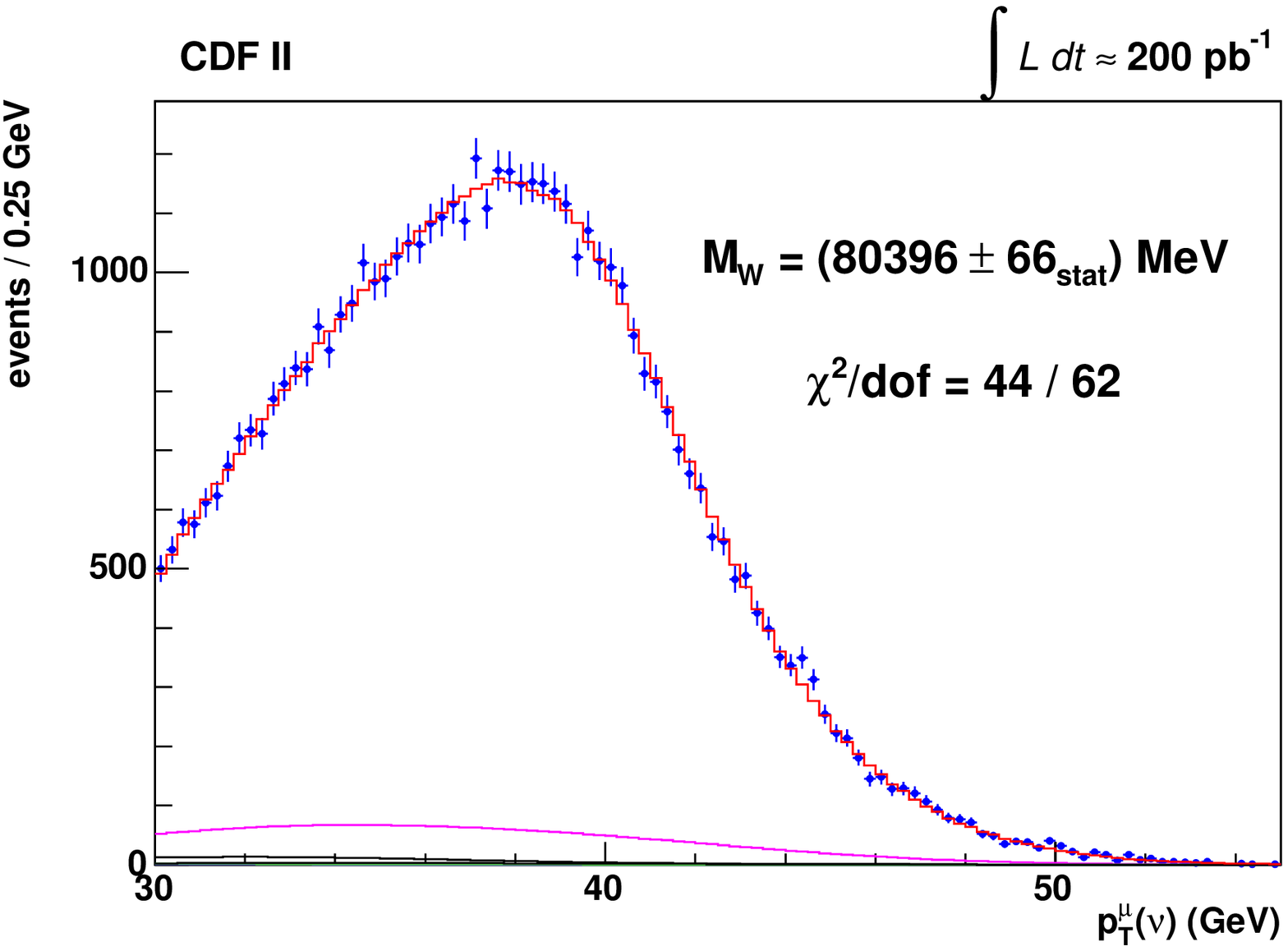}}
\end{center}
\vspace{-1pc}
  \caption{
Distributions of $M_W$ observables in CDF measurement (muon channel).
Blue -- data. Red -- fast simulation. Fit results and
statistical errors are indicated.
Left: $m_{T}$. Middle: muon $p_{T}$. 
Right: neutrino $p_{T}$.}
\label{fig:cdfmwfigmuon}
\end{figure*}

$M_W$ results from D0~\cite{d0mw} and CDF~\cite{cdfmw} along with
other $M_W$ measurements and combinations are shown inf Fig.~\ref{fig:mwfig}.
D0 result 80.401 $\pm$ 0.021(stat) $\pm$ 0.038(syst) GeV = 80.401 $\pm$ 0.043 GeV
agrees with the world average and the individual measurements and
is more precise than any other $M_W$ measurement from a single experiment.
CDF result 80.413 $\pm$ 0.034(stat) $\pm$ 0.034(syst) GeV = 80.413 $\pm$ 0.048 GeV.
Fig.~\ref{fig:mwd0} shows a comparison of observables between D0 1fb$^{-1}$ $W \rightarrow e \nu$  data
and fast simulation. 
Fig.~\ref{fig:cdfmwfigmuon} shows corresponding muon channel plots for CDF 0.2 fb$^{-1}$ measurement.
In both CDF and D0 measurements dominant experimental systematic error is due to lepton energy scale,
whereas dominant theoretical error is due to PDFs.
Currently both CDF and D0 experiments performed W boson mass measurements
only on small fraction of their data. These measurement lead to 31 MeV W mass
uncertainty from Tevatron and to world average uncertainty of 23 MeV.

Based on electroweak fits most probable Higgs mass value is 92 GeV, mass
region above 161 GeV is excluded at 95\% confidence level.
If world average uncertainty is reduced to 15 MeV then most probable value
and exclusion limit would become 71 GeV and 117 GeV respectively~\cite{peterenton}.
These estimates are made with the assumptions of no change in the central
values of W boson mass and top quark mass and with top quark mass uncertainty of 1 GeV.
With the full Tevatron dataset precision of 15 MeV may be possible~\cite{ashutoshandjan}.

\section{$W$ boson measurement at the LHC}
\label{sec:5}
$W$ mass measurements at the LHC is expected to involve additional challenges
in comparison with the corresponding measurements at the Tevatron.
First, much higher number of additional interactions, which produce in the detector
large energy deposits, uncorrelated with the $W$ boson.
Second, specifics of the $W$ boson production mechanisms.

In case of proton-antiproton collisions $W^+(W^-)$ is produced with valence $u$ and $\bar{d}$
($d$ and $\bar{u}$) quarks. Total number of produced $W^+$ and $W^-$ is the same.
As the $u$ quark tends to carry a higher fraction of the proton's 
momentum than the $d$ quark, the $W^+(W^-)$ is boosted, on average, 
in the proton(anti-proton) direction. 
Hence asymmetry in the production rate between $W^+$ and $W^-$ as a function of $W$ rapidity is observed.
However, $W$ boson $p_T$ spectrum, integrated over all rapidities is identical for $W^+$ and $W^-$.
in case of proton-antiproton collisions.

\begin{figure}[hbpt]
\begin{center}
\resizebox{0.85\columnwidth}{!}{ \includegraphics{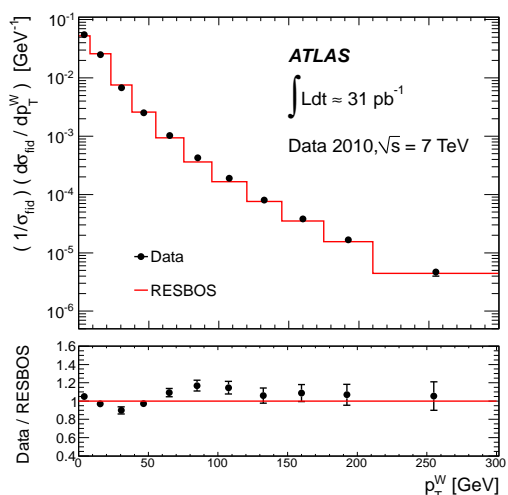} }
\end{center}
\vspace{-1pc}
\caption{Normalized differential cross section 
as a function of $W$ boson $p_T$ 
obtained from the combined electron and muon measurements, compared to the RESBOS prediction. 
}
\label{fig:atlaswpt}
\end{figure}

In case of proton-proton collisions $W^+(W^-)$ 
is produced with valence $u$ and sea $\bar{d}$ quarks
(valence $d$ and sea $\bar{u}$) quarks. Qualitatively the same type of asymmetry as
a function of $W$ boson rapidity is expected
as in case of proton-antiproton collisions. However the shape is expected to differ
compared to proton-antiproton collisions
since valence quarks and sea quarks have different momentum fraction distributions.
Besides the total number of produced $W^+$ is expected to exceed that of $W^-$
since proton contains two valence $u$ quarks and one valence $d$ quark.
The inclusive ratio of cross sections for $W^+$ and $W^-$ boson production
was measured by CMS to be 1.43$\pm$0.05~\cite{cmswasymold}.
Moreover, the shapes of  $W^+$ and $W^-$ $p_T$ spectra are expected to be different.
Hence, they need to be measured and modeled separately. Recently ATLAS measured 
$W$ boson $p_T$ spectrum \cite{atlaswpt}, shown in Fig.~\ref{fig:atlaswpt}, which can be considered first step towards
understanding this observable at the precision needed for $W$ mass measurement.
Another input needed for $W$ mass measurement is precise knowledge of parton distribution functions (PDFs).
Since $W$ charge asymmetry as a function of rapidity is driven by parton distributions, by measuring the asymmetry
parton distribution functions can be constrained. 
Asymmetry in the $W$ boson rapidity distribution
has traditionally been studied in terms of 
charged lepton asymmetry, as $W$ boson rapidity cannot be determined
on the event-by-event basis, since neutrino escapes the detection.
Charged lepton asymmetry, is the convolution of $W^{\pm}$ production 
and V-A (vector-axial vector) decay asymmetries.
Asymmetry is defined as a ratio of difference and sum of positively charged and negatively charged leptons.
ATLAS~\cite{atlas_det}, CMS~\cite{cms_det} and  LHCb~\cite{lhcb_det} already performed first measurements of lepton charge asymmetries \cite{lhcbswasym,atlaswasym,cmswasym}

Since sea quarks are involved in $W$ boson production at the LHC, both charm and strange quarks, unlike at the Tevatron,
contribute significantly. PDFs of both charm and strange quarks are currently very poorly constrained.
Significant improvements in PDF precision would be needed for precise $W$ mass measurement at the LHC.

\end{document}